# Mechanical configurable nanopatterning of polar topological states and formation of "post-skyrmion"

Lizhe Hu[1], Linming Zhou[1], Yuhui Huang[1], Sujit Das[2], He Tian[3], Yongjun Wu[1, 4, *], Zijian Hong[1, 4, *]

[1] School of Materials Science and Engineering, Zhejiang University, Hangzhou 310027, China

[2] Materials Research Centre, Indian Institute of Science, Bangalore 560012, India

[3] Center of Electron Microscopy, School of Materials Science and Engineering, Zhejiang University, Hangzhou, 310027, China

[4] Cyrus Tang Center for Sensor Materials and Applications, State Key Laboratory of Silicon Materials, Zhejiang University, Hangzhou, Zhejiang 310027, China

*Corresponding authors: Y.W. (yongjunwu@zju.edu.cn), Z.H. (hongzijian100@zju.edu.cn)

**Abstract**

The controllable phase transition and nanopatterning of topological states in a ferroelectric system under external stimuli are critical for realizing the potential applications in nanoelectronic devices such as logic, memory, race-track, etc. Herein, using the phase-field simulations, we demonstrate the mechanical manipulation of polar skyrmions in ferroelectric superlattices by applying external local compressive stress through an atomic force microscopy (AFM) tip. Different switching pathways are observed: under small to moderate force (<1 μN), the skyrmions coalesce to form a long stripe; while increasing the applied load (e.g., above 2 μN) leads to the suppression of spontaneous polarization, forming a new metastable topological structure, namely the "post-skyrmion". It is constructed by attaching multiple merons onto a center Bloch skyrmion, showing a topological charge of 1.5 (under 2 μN) or 2 (under 3 μN). We have further designed a mechanical nanopatterning process, where the stripes can form a designed pattern by moving the AFM tip (write), which can also be switched back to a full skyrmion state under an applied electric field (erase). We believe this study will spur further interest in mechanical manipulation and nanopatterning of polar topological phases through mechanical forces.

**Introduction**

Novel polar topological structures [1-3], including flux closure [4-5], vortex [6-11], skyrmion [12-14], polar wave [15, 16], hopfion [17], meron [18], and polar ice [19] etc., have been extensively investigated in functional oxide heterostructures (e. g. in low dimensional oxide superlattices). They are the host for various exotic physical properties such as chirality [20-22], negative capacitance [23-25], etc., and are envisioned for practical applications in next-generation high-density nonvolatile information storage and other nanoelectronic devices [6, 26, 27]. Moreover, the mutual conversions and transitions between different topological structures or regular domains have been demonstrated, by applying external electrical/mechanical/thermal/optical stimuli [1].

In particular, polar skyrmion, a whirl-like electric dipole configuration with a topological charge of 1 (akin to the ferromagnetic counterpart), has been observed in the $PbTiO_3/SrTiO_3$ (PTO/STO) heterostructures on a $SrTiO_3$ (STO) substrate [12]. It has been shown to exhibit topological protection where the topological charge is preserved under a small to medium electric field [24], with the shrinking or expansion of the skyrmion bubbles. The electric field manipulation of polar skyrmions has shown interesting physical phenomena such as the asymmetric distortion of the polar bubble, local topological transition before polar structural transition, and tunable negative capacitance [28]. Compared to the electrical switching with an applied electric field, mechanical switching through an atomic force microscopy (AFM) tip has shown great promise for device applications since it can avoid the field-driven problems such as charge injection and dielectric breakdown [29]. Previously, mechanical switching of the topological structures such as polar vortex has been investigated both experimentally and theoretically [30-33]. It is discovered that the polar vortices or flux-closure domains can be reversibly switched to regular domains (*a*-domains/*c*-domains) through nanoindentation [30-31]. Whereas the chirality of the polar vortex can be mechanically switched by applying a torsion force [32] or through careful design of the geometry of the ferroelectric materials [33]. Meanwhile, deterministic mechanical switching of the polar skyrmions has yet to be discovered.

Herein, we employ the phase-field method to investigate the kinetic evolution of polar skyrmions under the applied stress through an AFM tip. With small to medium applied stress (e.g., <1 μN), the neighboring skyrmions merged to form a stripe. While under relatively large stress (above 2 μN), the polarization is greatly suppressed, forming a new metastable topological structure, namely the "post-skyrmion". It is composed of a Bloch skyrmion and several merons, yielding a topological charge of 1.5 (under 2 μN load) or 2 (under 3 μN load). The mechanical writing of nanopatterns is further demonstrated, where the long stripe can be manipulated by moving the AFM tip along the designed pathway. The as-written structure can be fully erased with an externally applied electric field. This study paves the way for the mechanical manipulation and nanopatterning of the polar topological structures in oxide heterostructures.

**Main**

Details of the phase-field method are given in Methods and previous reports [8, 9, 34-37]. All the simulation parameters are taken from the literature [38-40]. The simulation setup is given in **Fig. 1**. Fig. 1(a) shows the schematics of the simulation system, where periodic stackings of (PTO)$_{16}$/(STO)$_{16}$ layers are grown on an STO substrate. An AFM tip is located on top of the film to apply local external stress. The planar view of the initial polar pattern without external stress is shown in Fig. 1(b)-1(c), which demonstrates the formation of circular bubble-like structures with center divergent in-plane polarizations, consistent with previous reports [12, 24]. Further Pontryagin density calculations confirm that the bubbles are polar skyrmions with a defined topological charge of 1 (**Fig. S1**). The schematic diagram of the contact structure is illustrated in Fig. 1(d), the radius of the tip and contact area is $R$ and $a$, respectively. The applied stress distribution under the AFM tip can be described by a spherical indenter [41]: $\sigma_{33} = -\frac{3p}{2\pi a^2}\sqrt{1-\frac{r^2}{a^2}}$, where $p$ is the applied load, $r$ is the distance to the tip center. The initial stress distributions without external force in the planar view of the top PTO layer and cross-section view of the superlattice are plotted in **Fig. S2**. It can be seen that the skyrmion walls show tensile stress with in-plane polarizations while the other *c*-like regions are under compressive stress. The stress distribution after applying a 1 μN load is plotted in Fig. 1(e), where a circular dark blue region is observed in the planar view, showing the compressive applied stress through the AFM tip. While the cross-sectional view demonstrates that the influence of the externally applied stress is mainly concentrated on the top STO and PTO layers, and decays quickly along the thickness direction (Fig.1f). This indicates that unlike the electric field applied through a PFM tip which could penetrate through the thickness of the whole film, the stress field with an AFM tip is more localized. This could facilitate the localized control of the polar pattern on a single PTO layer. The line plots of the local strain (Fig. S2c) and stress (Fig. S2d) with and without external pressure along the top line (the horizontal yellow dotted lines in Fig. 1f and Fig. S2a-S2b) demonstrate that the maximum stress under the tip is close to ~4 GPa, corresponding to a maximum strain increase of 2.4 % as compared to that without external force. The line plots of the local stress (Fig. S2e) with and without external pressure along the thickness direction are given to confirm the localized applied stress/strain field on the first PTO layer.

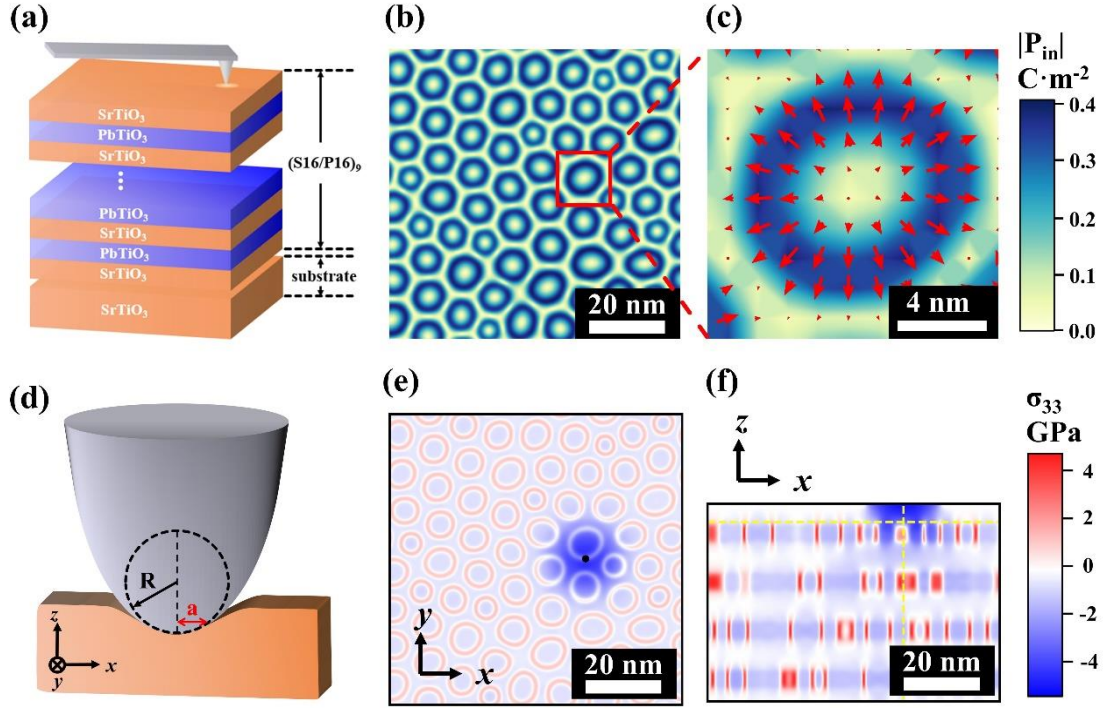

**Fig. 1| Initial setup for this study.** (a) Schematics of the PTO/STO superlattice system with the AFM tip to apply stress. (b) Planar view of the in-plane polarization magnitude from phase-field, showing the formation of skyrmion bubbles. (c) Magnified view of the polar skyrmion bubble marked with a red box in (b), overlaid with the in-plane polar vector. (d) The magnified schematic of the AFM tip part. $R$ is the radius of the AFM tip sphere and $a$ is the radius of the contact area. (e) Distribution of stress component $\sigma_{33}$ in the $x$-$y$ plane at the top of PTO layer under 1 μN tip load. The center black dot represents the center of the tip. (f) Distribution of stress component $\sigma_{33}$ in the cross-section of the top few STO and PTO layers under the 1 μN tip load.

The kinetic evolution of the skyrmions after applying external stress is shown in **Fig. 2**. A 1 μN load is applied on the skyrmion wall (the red dot in Fig. 2a) through an AFM tip with a tip radius of 10 nm. It can be seen that, after 1000 timesteps, the skyrmion wall is broadened with the reduction of polarization (Fig. 2b). The line plot of the local polarization evolution

(**Fig. S3**) confirms the decrease of both in-plane and out-of-plane polarization components. This can be understood since the applied out-of-plane force will suppress the lattice constant along the out-of-plane direction while the in-plane lattice constants expand, which ultimately decreases the tetragonality and hence the polarization of the PTO layers. Consequently, after 2000 timesteps, the merging of two neighboring skyrmions to form a single skyrmion is observed (Fig. 2c). Then, another neighboring skyrmion also coalesced, forming an L-like pattern underneath the tip after 6000 timesteps (Fig. 2d). Eventually, it transforms into an oval-like bubble structure to reduce the gradient and surface energies (Fig. 2e). Further, calculation of the Pontryagin density indicates that an oval-like bubble also has a topological charge of +1 (Fig. 2f), showing that the whole kinetic transition is a topological phase transformation which merges three skyrmions to a single oval skyrmion with the reduction of the total skyrmion number or topological charge. After removing the applied stress (**Fig. S4**), the newly formed oval skyrmion is stable, with the recovery of the polarization magnitude. Thus, the local mechanical manipulation of the polar skyrmion is achieved by applying 1 μN through an AFM tip, where several skyrmions can merge into one in the vicinity of the tip.

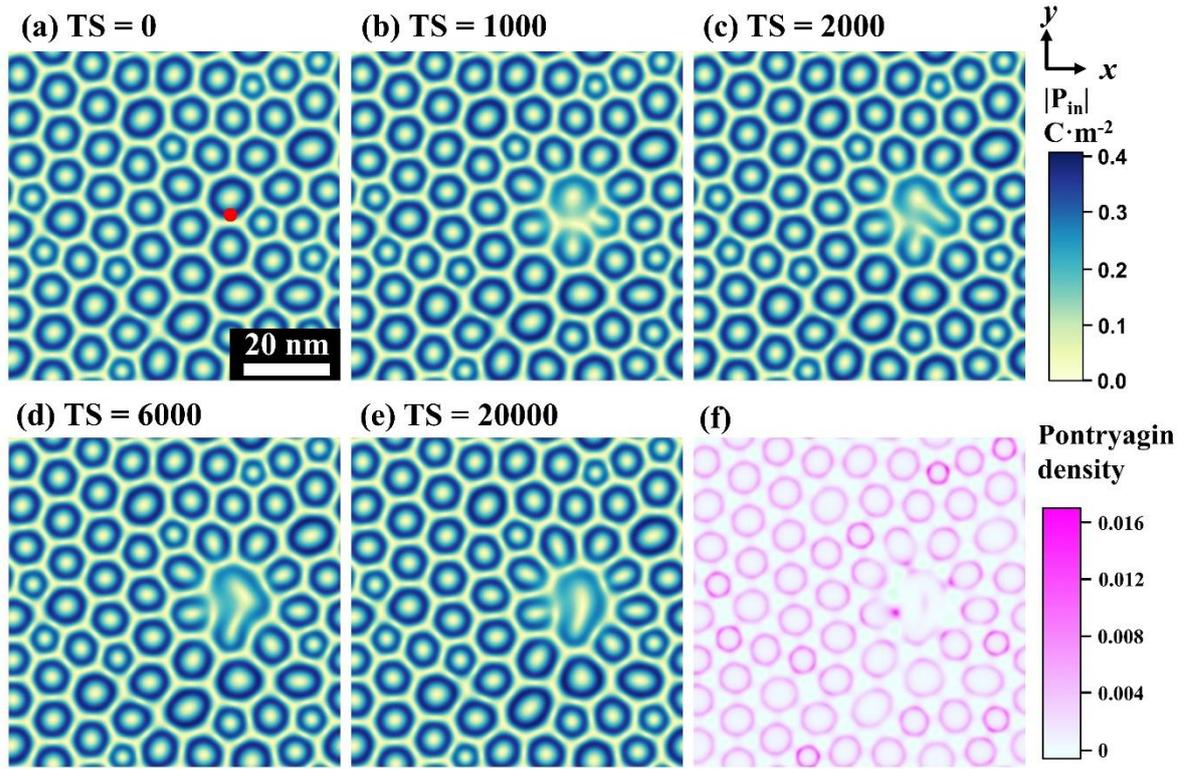

**Fig. 2| Kinetics evolution process for polar skyrmions under 1 μN tip load.** (a-e) The kinetic evolution pathway after 0, 1000, 2000, 6000, and 20000 timesteps, respectively. (f) Pontryagin density distribution on the top PTO layer after 20000 timesteps.

Next, we proceed to understand the influence of applied load magnitude on the skyrmion switching kinetics (**Fig. 3**), with applied force ranging from 0.5 μN to 3 μN. Under a small applied load, e.g., 0.5 μN (Fig. 3a), the polarization of the top PTO layer underneath a tip is slightly reduced while the overall skyrmion structure is preserved. Further increasing the load to 1 μN leads to a local topological phase transformation, where several small skyrmion bubbles have merged to form a single large skyrmion (Fig. 3b), as has been discussed before. Increasing the applied load to 2 μN, a sea-star-like structure (referred to as the "post-skyrmion" structure) can be discovered in the PTO layer just beneath the tip (Fig. 3c). The magnified view of the in-plane polar distribution for this novel structure is given in Fig. 3(d). It is observed that the "post-skyrmion" is composed of an in-plane vortex-like structure in conjunction with five "hooks" in the surroundings. The topological feature of the "post-skyrmion" is further calculated (Fig. 3e). The surface integration of the Pontryagin density shows that the central in-plane vortex-like structure has a topological charge of -1, while each "hook" has a topological charge of +0.5. This suggests that the in-plane vortex-like structure can be classified into "Bloch skyrmion", while each "hook" is a "meron" structure. The combination of a Bloch skyrmion and five merons gives a net topological charge of +1.5 for the "post-skyrmion". The distributions of different polarization components after evolution under different applied loads (Fig. 3f and **Fig. S5**) indicate the continuous reduction of the polarization magnitude with increasing the applied load.

To better understand the detailed formation process of the "post-skyrmion", the kinetic evolution pathway and corresponding topological features of the skyrmions under the 2 μN tip load are illustrated (**Fig. S6**). As shown in Fig. S6(a), after 1000 timesteps, the skyrmions underneath the AFM tip are "melted" due to the reduction of the spontaneous polarization. This is accompanied by the decrease of the total topological charge from 9 (the sum of the nine skyrmions) to 7.5 in the surrounding region (Fig. S6e). Then, after 2000 timesteps, the "melted" region expands to form a meron-like structure (Fig. S6b), with a further decrease of the topological charge to 5.5 (Fig. S6f). Consequently, after switching for 6000 timesteps (Fig. S6c), the polarizations surrounding the tip center have been rearranged into an in-plane vortex-like (Bloch skyrmion) structure, while merons are attached to the Bloch skyrmion to form a sea-star-like structure, with one side of the merons connected to the Bloch skyrmion and the other side is open. Eventually, after 20000 timesteps (Fig. S6d), a small meron on the left side disappears completely, with the expansion of the neighboring skyrmion. While another skyrmion in the vicinity of the tip is "melted" into a meron and attached to the center Bloch skyrmion, leading to the formation of the "post-skyrmion". At this stage, the total topological charge of the sea-star-like structure evolves to +1.5 (Fig. S6g-h).

Further increasing the applied load to 3 µN, a similar "post-skyrmion" structure is observed (Fig. 3g). While notably, the size of the "post-skyrmion" increases, which could attach more merons onto the Bloch skyrmion. This can be understood since the larger the applied force, the larger the contact area for the tip. In this case, the "post-skyrmion" is composed of six merons and a Bloch skyrmion, showing a net topological charge of 2. The line plot of the out-of-plane polarization distribution (Fig. 3f) shows the continuous decrease of the polarizations near the AFM tip and the expansion of the switched region with increasing applied load.

The polar structures after the removal of the applied stresses are highlighted in Fig. 3(h)-3(k). When the applied load is small (e.g., 0.5 µN), after the applied stress is removed, it recovers to the original state. For the 1 µN case, as analyzed above, an oval-like skyrmion is obtained in the system even after the removal of the applied load. While the "post-skyrmion" formed under high applied stress (e. g., 2 µN and 3 µN) is no longer retained, which decomposes to skyrmion and stripe domains, indicating that the "post-skyrmion" is a metastable state.

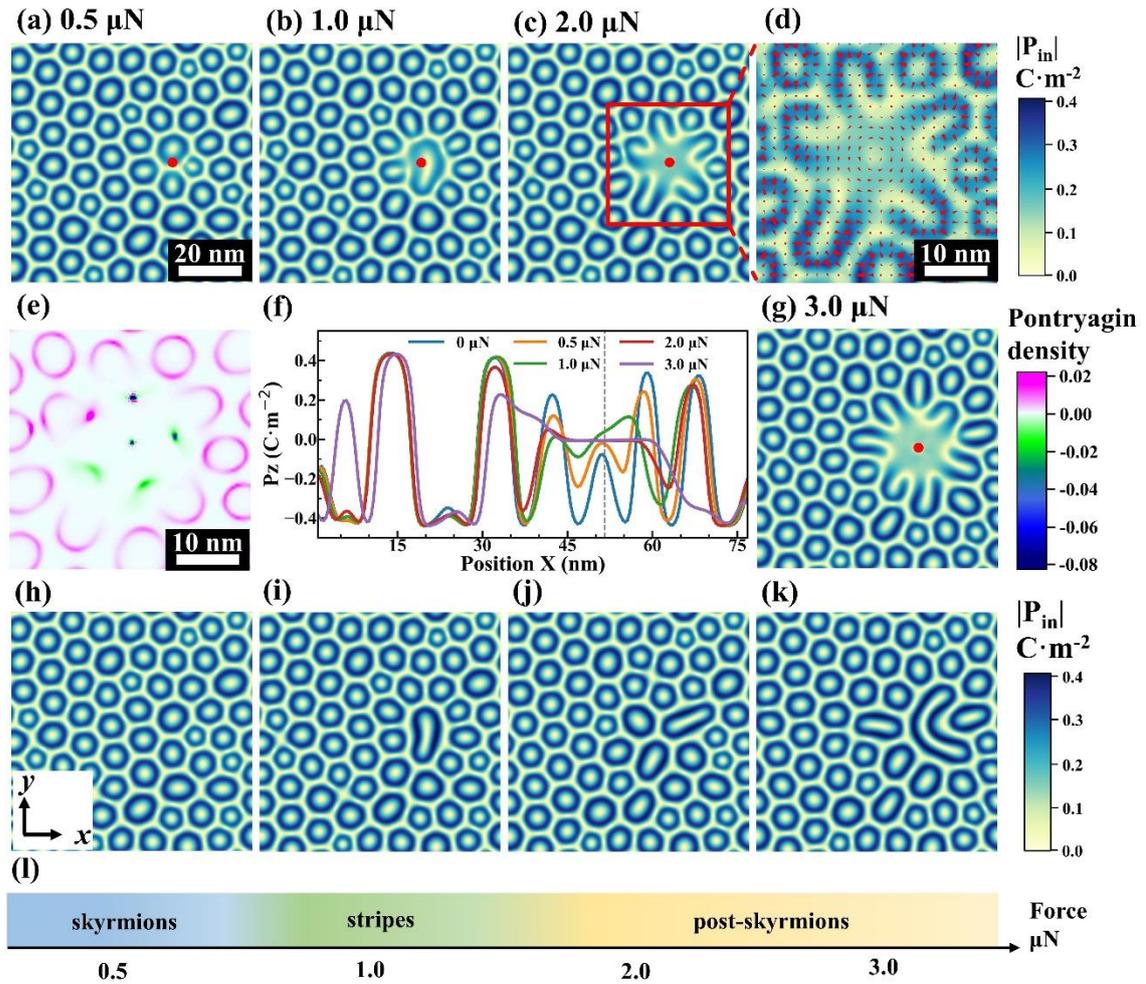

**Fig. 3| Topological evolution patterns under different applied loads.** (a-c) The in-plane polarization distribution of the top PTO layer under the load of 0.5, 1.0, and 2.0 μN, respectively. (d) Magnified view of the "post-skyrmion" overlaid with the in-plane polar vector. (e) Magnified view of the Pontryagin density distribution of the "post-skyrmion". (f) Distribution of out-of-plane polarization under different loads. (g) The in-plane polarization distribution of the top PTO layer under the load of 3.0 μN. (h-k) The in-plane polarization distribution of the top PTO layer after removal of the 0.5, 1.0, 2.0, and 3.0 μN load, respectively.

Having understood the mechanical switching kinetics with a static AFM tip, now we proceed to design a writing pathway by moving the tip to facilitate the localized "writing" of a specific pattern (**Fig. 4**). First, we randomly select a position from the upper right of the simulation box (Fig.4a), and apply a 1 μN load with an AFM tip, then the tip is removed and the system relaxes to equilibrium. A stripe domain is obtained, as shown in Fig.4 (b). Consequently, the AFM tip moves along the in-plane *y* direction through a designed route. The "merging" of more skyrmions to form an elongated stripe along the tip moving direction is discovered (Fig. 4c). Interestingly, the stripe can turn 90° with the changing of the tip moving direction. Eventually, after completing 13 "apply load"-"remove load" cycles, a "box" has been successfully obtained as designed (Fig. 4g). In the last cycle, the written circular stripe meets at both ends, changing from one long strip to two nested skyrmions where eight small skyrmions are enclosed inside. To unravel the "box" formed by the writing process, a uniform electric field of 900 kV/cm is applied through a plate electrode. Under the applied electric field, the skyrmions begin to shrink, while the long stripe decomposes into small bubbles (Fig. 4h) and returned to the regular pattern after the removal of the field. Overall, in this set of experiments, we have realized the writing of nanopattern through an AFM tip by designing the tip moving direction. While the nanopattern can be fully erased by an applied capacitor electric field.

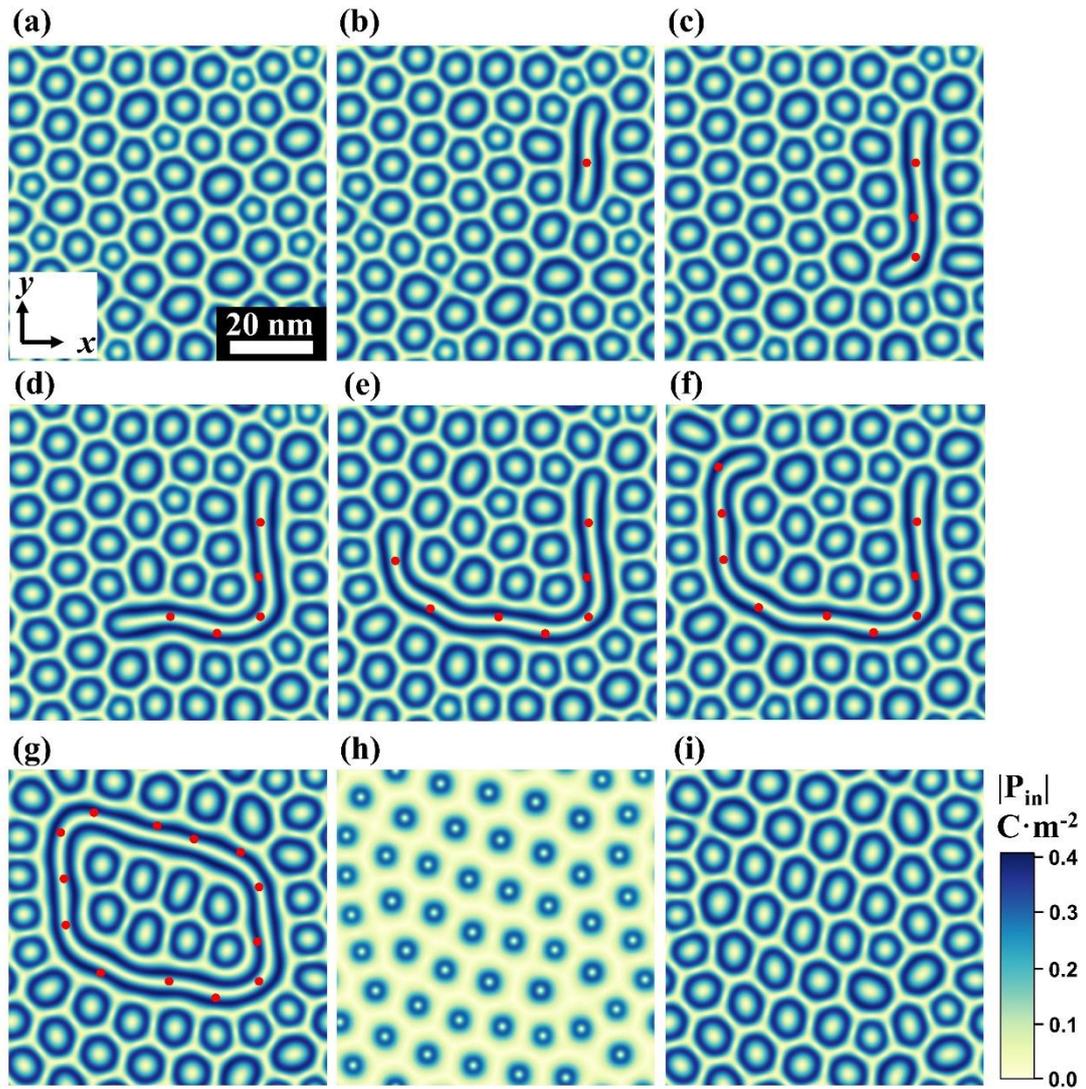

**Fig. 4| Mechanical nanopatterning using an AFM tip**. (a-g) Process of writing locally in the superlattice system after 0, 1, 3, 5, 7, 9, and 13 "apply-remove load" cycles, respectively. (h) Erasing the writing box and the evolution of the system after applying 900 kV/cm uniform electric field on (g). (i) Regular skyrmion pattern after removing the field.

**Conclusion**

In conclusion, we report the mechanical manipulation and nanopatterning of polar skyrmions through an AFM tip in the PTO/STO superlattice system. Different switching patterns are observed, where under a relatively low applied pressure of 1 μN, several skyrmion bubbles around the tip coalesce to form a skyrmion stripe that does not revert to bubbles after the stress is removed. While with a larger applied load (such as 2 μN or 3 μN), the polarization underneath the tip is greatly suppressed. A "post-skyrmion" structure is formed, which is composed of a Bloch skyrmion in the center and multiple merons, yielding a net topological charge of 1.5 (under 2 μN ) or 2 (under 3 μN ). The "post-skyrmion" is metastable and decomposes into skyrmions and stripes when the externally applied load is removed. A mechanical nanopatterning process is demonstrated, where the stripe can move along the tip moving direction, leading to the writing of a stable desired pattern. Moreover, this writing process is reversible where the formed pattern can be fully erased with an applied electric field which recovers to the initial full skyrmion state when the field is removed.

Our findings demonstrate the possibility to form new topological states by coalescence of existing topological states, indicating the potential design of "topological computing", where the topological charge can be utilized as a computing bit. These observations also bring a key question for future exploration: how can we stabilize a "post-skyrmion" without external force? This might be achieved through design of the mechanical boundary conditions. We hope to spur further theoretical and experimental interest in the mechanical manipulation and nanopatterning of the polar topological structures in oxide heterostructures.

**Methods**

*Phase-field simulations*. Phase-field simulations were utilized to simulate the evolutions of the PTO/STO superlattice, on a STO substrate. The spontaneous polarization $P_i$ was governed by the time-dependent Ginzburg–Landau (TDGL) equations [34-36]:

$$\frac{\partial P_i(\vec{r},t)}{\partial t} = -L\frac{\delta F}{\delta P_i(\vec{r},t)}, (i = 1, 2, 3)$$

where $t$ and $L$ are the evolution time and kinetic coefficient, respectively. The polarization field is represented by $P_i$. The free energy functional is represented by $F$, and can be expressed as the volume integration of the Landau, mechanical, electrical, and gradient energy densities:

$$F = \int (f_{Landau} + f_{elastic} + f_{electric} + f_{gradient})dV$$

Detailed expressions of the energy densities, the numerical calculations, the phase-field equations, and the simulation parameters can be found in previous reports [8, 9, 34, 37-40]. A 3D mesh of 192 × 192 × 350 was used, with a grid spacing of 0.4 nm. Periodic boundary conditions were applied on the two in-plane dimensions, while a superposition method [41] is used on the out-of-plane direction. The thickness direction consists of 30 grids of the substrate, 300 grids of thin films with periodic stacking of $(PTO)_{16}/(STO)_{16}$, and 20 layers of air. The normalized timestep is set as 0.01 in this study.

**Acknowledgement**

This work is supported by the Joint Funds of the National Natural Science Foundation of China (grant no. U21A2067, YW) and the Fundamental Research Funds for the Central Universities (No. 2021FZZX003-02-03, ZH). ZH gratefully acknowledges a start-up grant from Zhejiang University. SD gratefully acknowledges a start-up grant from the Indian Institute of Science, Bangalore, India. The phase-field simulation was performed on the MoFang III cluster on Shanghai Supercomputing Center (SSC).

**Declare of Interest**

The authors declare no conflict of interest.

**Data Availability Statement**

The data that support the findings of this study are available from the corresponding author upon reasonable request.

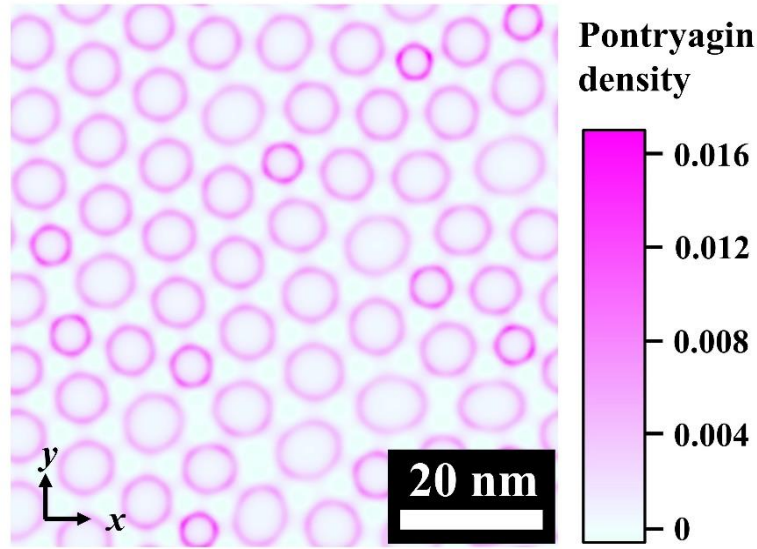

**Fig. S1| Pontryagin density distribution for the initial bubble structure.**

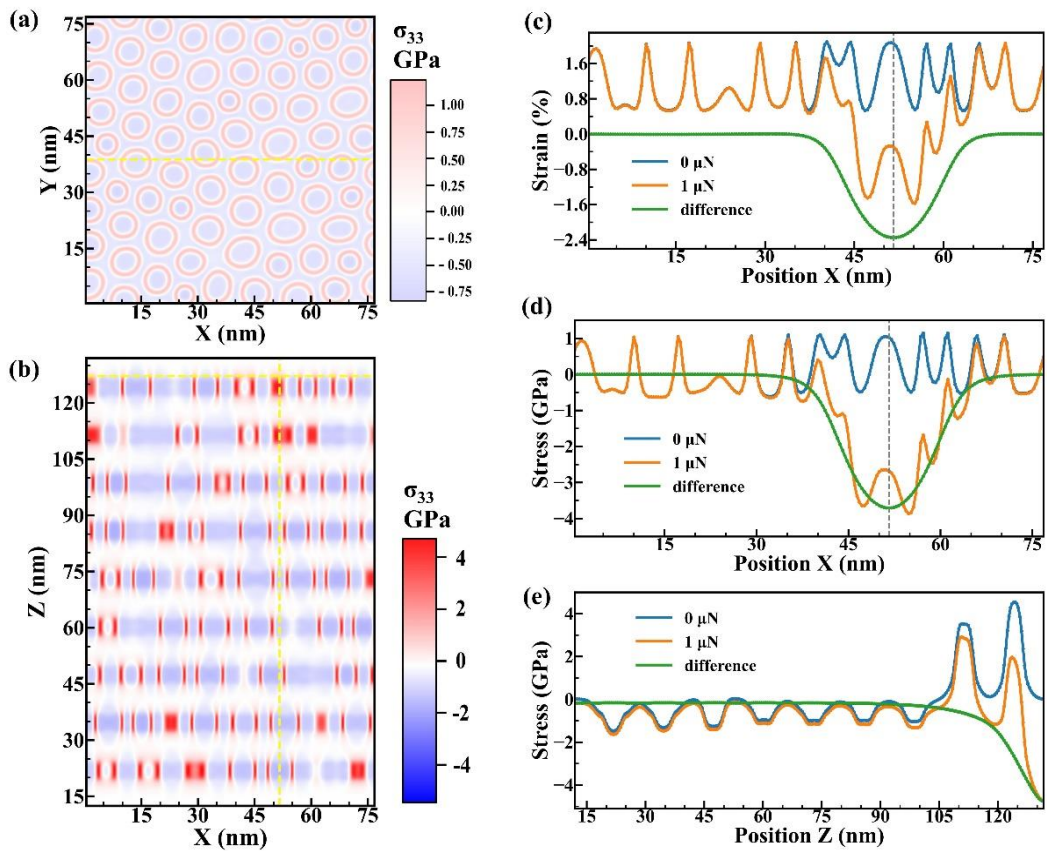

**Fig. S2| Initial stress distributions without external forces.** (a) Distribution of stress component $\sigma_{33}$ in the *x-y* plane at the top of PTO layer without extra stress. (f) Distribution of

stress component σ₃₃ in the cross-section without extra stress. (c) and (d) The comparison of strain(c) and stress(d) with and without external pressure along the top line (the horizontal yellow dotted lines in Fig. 1f and Fig. S2a-S2b) under the tip at tip PTO layer. (e) The line plot of the stress (Fig. S2e) with and without external pressure along the thickness direction under the tip (the vertical yellow dotted lines in Fig. 1f and Fig. S2b).

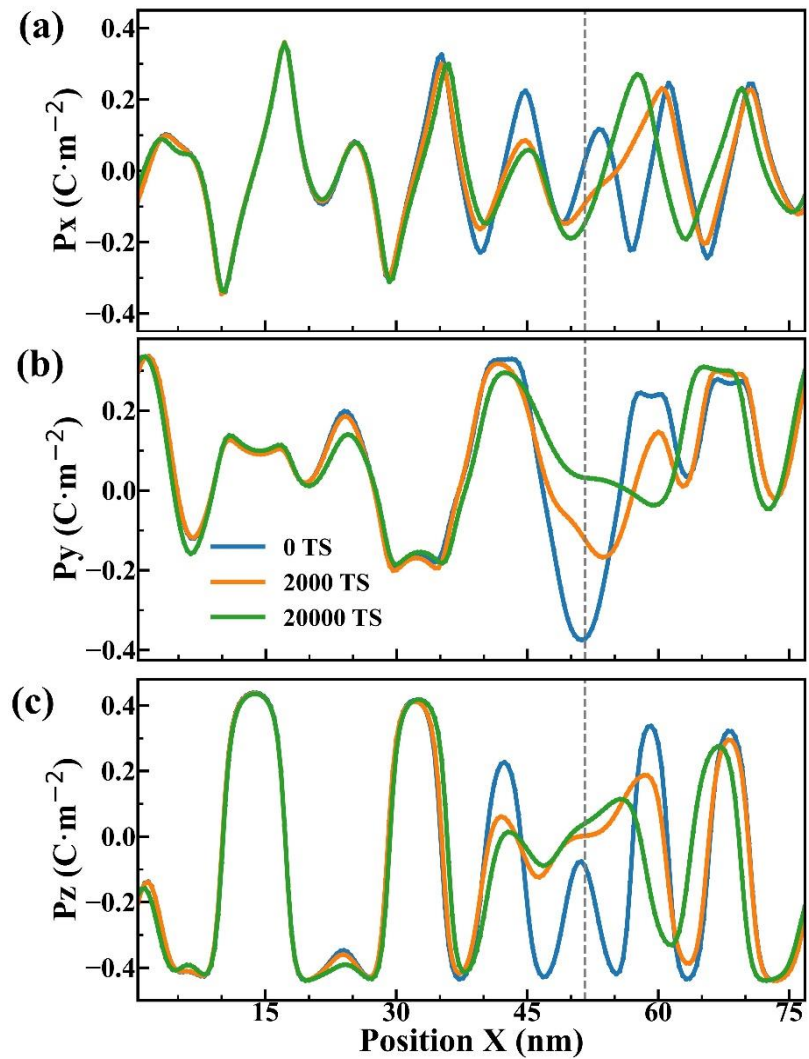

**Fig. S3| Line plots of the polarization distributions after applying a 1 μN tip load.** (a-c) Distribution and temporal evolution of $P_x$, $P_y$, and $P_z$, respectively.

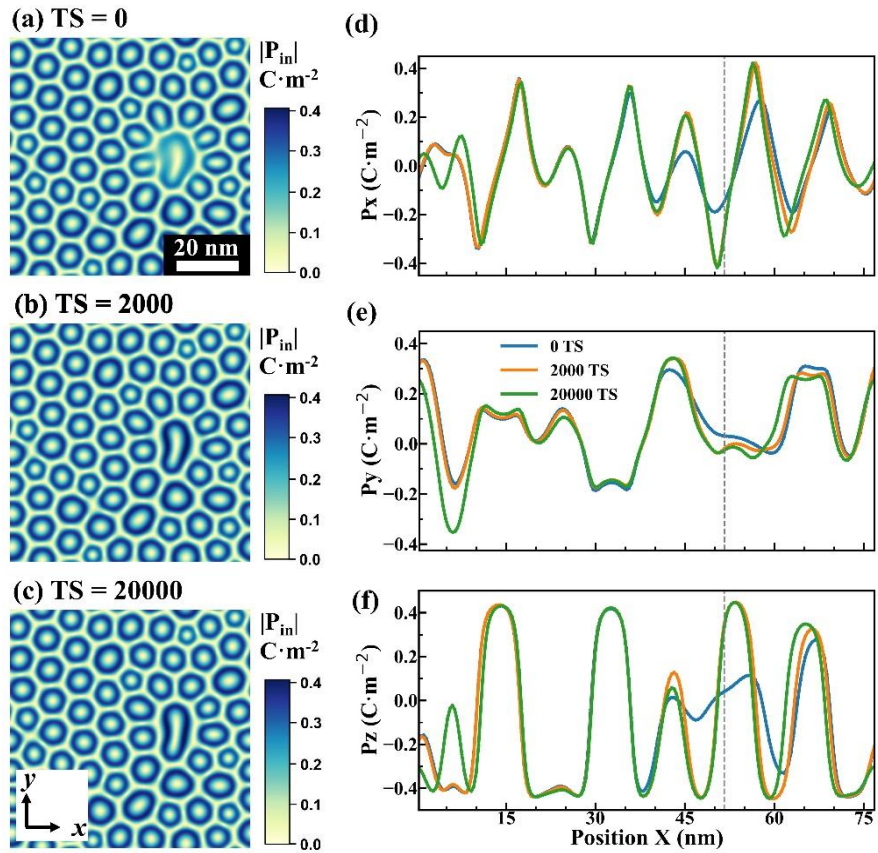

**Fig. S4| Kinetics of the skyrmion evolution process after the extra stress is removed.** (a-c) The kinetics evolution pathway after 0, 2000, and 20000 timesteps, respectively. (d-f) Distribution and evolution of $P_x$, $P_y$, and $P_z$ after removal of the 1 μN load, respectively.

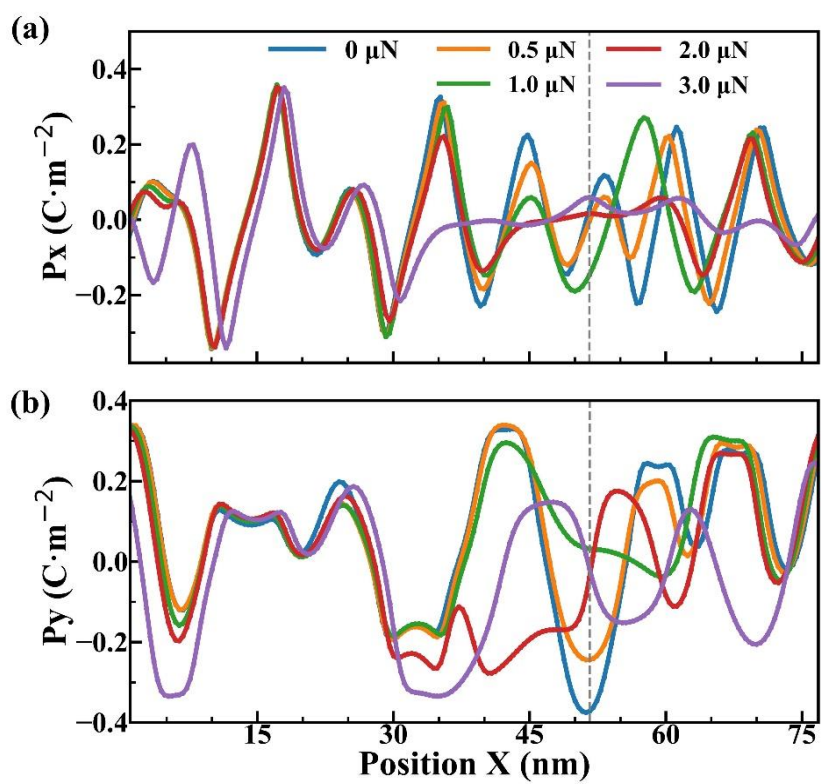

**Fig. S5| Line plots of the polarizations after applying different loads.** (a-b) Distribution of $P_x$ and $P_y$ under different forces, respectively.

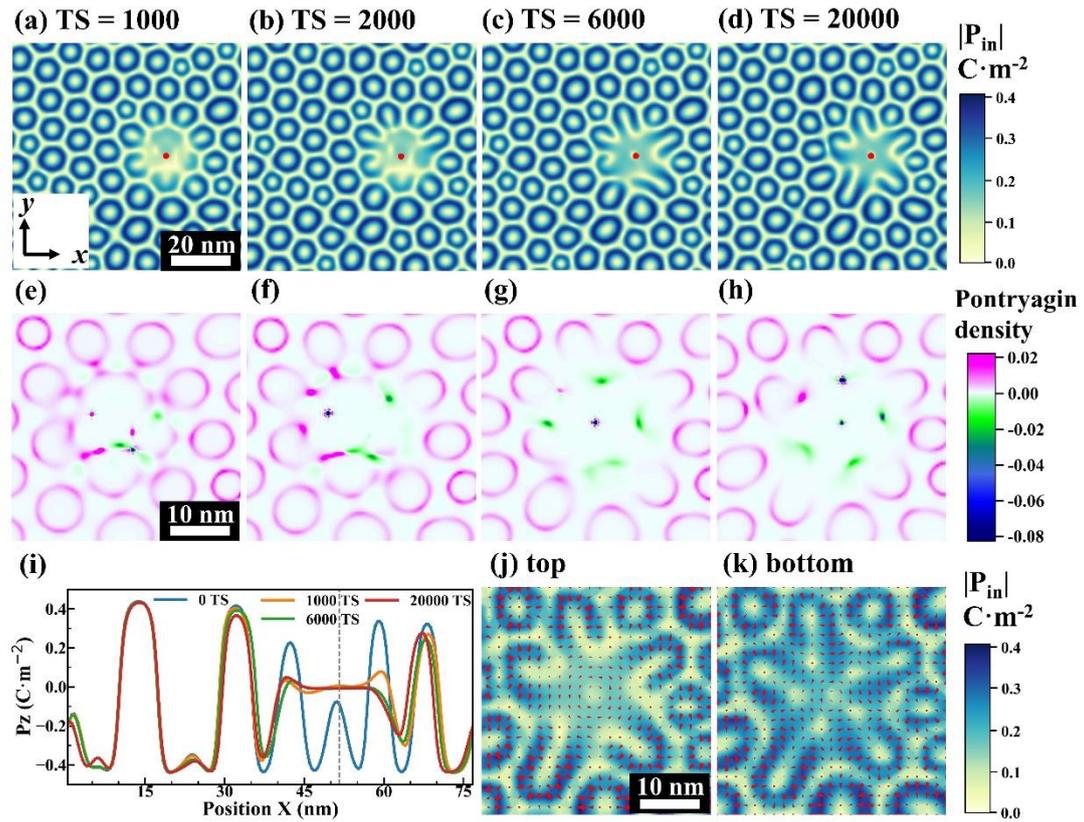

**Fig. S6| Kinetics and topological transition of the skyrmions under 2 µN tip load.** (a-d) The kinetic evolution pathway after 1000, 2000, 6000, and 20000 timesteps, respectively. (e-h) corresponding topological features of (a-d) respectively. (i) Distribution and evolution of out-of-plane polarization. (j-k) The topmost and the bottommost in-plane polarization distributions of the "post-skyrmion" within the PTO layer.